\def\etal{{\it et al.}}
\def\ie{{\it i.e.}}

\def\~{{$\tilde{\phantom{a}}$}}

\documentclass [12pt] {article}
\usepackage{epsfig}
\usepackage{color}

\def\thebibliography#1{\section{References}\markboth
 {REFERENCES}{REFERENCES}\list
 {[\arabic{enumi}]}{\settowidth\labelwidth{[#1]}\leftmargin\labelwidth
 \advance\leftmargin\labelsep
 \usecounter{enumi}}
 \def\newblock{\hskip .11em plus .33em minus -.07em}
 \sloppy
 \sfcode`\.=1000\relax}
\def\upcite#1{\raise6pt\hbox{\scriptsize
\cite{#1}}}
\def\lsim{\mathrel {\vcenter {\baselineskip 0pt \kern 0pt
    \hbox{$<$} \kern 0pt \hbox{$\sim$} }}}
\def\gsim{\mathrel {\vcenter {\baselineskip 0pt \kern 0pt
    \hbox{$>$} \kern 0pt \hbox{$\sim$} }}}
\def\gtlt{\mathrel {\vcenter {\baselineskip 0pt \kern 0pt
    \hbox{$>$} \kern 0pt \hbox{$<$} }}}

\def\hline{\noalign{\hrule \vskip2pt}}

\def\|{\ifmmode\Vert\else \char`\|\fi}
\ifx\oldzeta\undefined                          
  \let\oldzeta=\zeta                            
  \def\zzeta{{\raise 2pt\hbox{$\oldzeta$}}}     
  \let\zeta=\zzeta                              
\fi

\ifx\oldchi\undefined                           
  \let\oldchi=\chi                              
  \def\cchi{{\raise 2pt\hbox{$\oldchi$}}}       
  \let\chi=\cchi                                
\fi

\def\frac#1#2{{#1 \over #2}}

\def\half{\ifinner {\scriptstyle {1 \over 2}}
   \else {1 \over 2} \fi}

\def\ave#1{\left\langle#1\right\rangle} 

\def\abs#1{\left\vert#1\right\vert}	

\def\simge{\mathrel{%
   \rlap{\raise 0.511ex \hbox{$>$}}{\lower 0.511ex \hbox{$\sim$}}}}
\def\simle{\mathrel{
   \rlap{\raise 0.511ex \hbox{$<$}}{\lower 0.511ex \hbox{$\sim$}}}}

\def\buildchar#1#2#3{{\null\!                   
   \mathop#1\limits^{#2}_{#3}                   
   \!\null}}                                    
\def\overcirc#1{\buildchar{#1}{\circ}{}}

\def\slashchar#1{\setbox0=\hbox{$#1$}           
   \dimen0=\wd0                                 
   \setbox1=\hbox{/} \dimen1=\wd1               
   \ifdim\dimen0>\dimen1                        
      \rlap{\hbox to \dimen0{\hfil/\hfil}}      
      #1                                        
   \else                                        
      \rlap{\hbox to \dimen1{\hfil$#1$\hfil}}   
      /                                         
   \fi}                                         %

\def\subrightarrow#1{
  \setbox0=\hbox{
    $\displaystyle\mathop{}
    \limits_{#1}$}
  \dimen0=\wd0
  \advance \dimen0 by .5em
  \mathrel{
    \mathop{\hbox to \dimen0{\rightarrowfill}}
       \limits_{#1}}}                           

\def\overlay#1#2{\ifmmode%
\setbox0=\hbox{$#1$}%
\setbox1=\hbox to\wd0{\hss$#2$\hss}\else%
\setbox0=\hbox{#1}%
\setbox1=\hbox to\wd0{\hss#2\hss}\fi%
#1\hskip-\wd0\box1 }

\def\pmb#1{\leavevmode\setbox0=\hbox{#1}%
\kern-.02em\copy0\kern-\wd0
\kern.04em\copy0\kern-\wd0
\kern-.02em\raise.04em\box0 }

\def\vereq#1#2{\lower3pt\vbox{\baselineskip1.5pt \lineskip1.5pt
\ialign{$\m@th#1\hfill##\hfil$\crcr#2\crcr\sim\crcr}}}

\def\tensor#1{\protect\@ontopof{#1}{\leftrightarrow}{1.15}\mathord{\box2}}
\def\overstar#1{\protect\@ontopof{#1}{\ast}{1.15}\mathord{\box2}}
\def\overdots#1{\protect\@ontopof{#1}{\cdots}{1.0}\mathord{\box2}}
\def\overcirc#1{\protect\@ontopof{#1}{\circ}{1.2}\mathord{\box2}}
\def\loarrow#1{\protect\@ontopof{#1}{\leftarrow}{1.15}\mathord{\box2}}
\def\roarrow#1{\protect\@ontopof{#1}{\rightarrow}{1.15}\mathord{\box2}}

\def\@ontopof#1#2#3{%
{\mathchoice
{\@@ontopof{#1}{#2}{#3}\displaystyle\scriptstyle}%
{\@@ontopof{#1}{#2}{#3}\textstyle\scriptstyle}%
{\@@ontopof{#1}{#2}{#3}\scriptstyle\scriptscriptstyle}%
{\@@ontopof{#1}{#2}{#3}\scriptscriptstyle\scriptscriptstyle}%
}%
}

\def\@@ontopof#1#2#3#4#5{%
\setbox0=\hbox{$#4#1$}%
\setbox1=\hbox{$#5#2$}%
\setbox2=\hbox{}\ht2=\ht0 \dp2=\dp0 %
\ifdim\wd0>\wd1 %
\setbox1=\hbox to\wd0{\hss\box1\hss}%
\mathord{\rlap{\raise#3\ht0\box1}\box0}%
\else   %
\setbox1=\hbox to.9\wd1{\hss\box1\hss}%
\setbox0=\hbox to\wd1{\hss$#4\relax#1$\hss}%
\mathord{\rlap{\copy0}\raise#3\ht0\box1}%
\fi
}%

\def\lambdabar{\protect\@lambdabar}
\def\@lambdabar{%
\relax
\bgroup
\def\@tempa{\hbox{\raise.73\ht0
\hbox to0pt{\kern.25\wd0\vrule width.5\wd0
height.1pt depth.1pt\hss}\box0}}%
\mathchoice{\setbox0\hbox{$\displaystyle\lambda$}\@tempa}%
{\setbox0\hbox{$\textstyle\lambda$}\@tempa}%
{\setbox0\hbox{$\scriptstyle\lambda$}\@tempa}%
{\setbox0\hbox{$\scriptscriptstyle\lambda$}\@tempa}%
\egroup
}

\def\corresponds{{\lower.2ex\hbox{=}}{\rm\kern-.75em^\triangle}}
\def\succsim{\succ\kern-.9em_\sim\kern.3em}
\def\precsim{\prec\kern-1em_\sim\kern.3em}
\def\slantfrac#1#2{\kern1em^{#1}\kern-.3em/\kern-.1em_{#2}}

\def\zzeta{{\small \zeta}}
\def\zzzeta{{\tiny \zeta}}

\begin{document}
                                                                
\begin{center}
{\Large\bf Gaussian Laser Beams via Oblate Spheroidal Waves}
\\

\medskip

Kirk T.~McDonald
\\
{\sl Joseph Henry Laboratories, Princeton University, Princeton, NJ 08544}
\\
(October 19, 2002)
\end{center}

\section{Problem}

Gaussian beams provide the simplest mathematical description of the
essential features of a focused optical beam, by ignoring higher-order
effects induced by apertures elsewhere in the system.

Wavefunctions $\psi({\bf x},t) = \psi({\bf x}) e^{-i \omega t}$
for Gaussian laser beams \cite{paraxial,Goubau,Boyd,Kogelnik,%
Lax,Davis,Barton,diff,axicon,bessel} 
of angular frequency $\omega$ are typically deduced in
the paraxial approximation, meaning that in the far zone the functions 
are accurate only for angles $\theta$ with respect to the beam axis that
are at most a few times the characteristic diffraction angle 
\begin{equation}
\theta_0 = {\lambda \over \pi w_0}
 = {2 \over k w_0}
= {w_0 \over z_0}\, ,
\label{p8}
\end{equation} 
where $\lambda$ is the wavelength,
$ k = \omega / c = 2 \pi / \lambda$ is the wave number, $c$ is the
speed of light, $w_0$ is the radius of the beam waist, and $z_0$ is the
depth of focus, also called the Rayleigh range, which is related by
\begin{equation}
z_0 = {k w_0^2 \over 2}
 = {2 \over k \theta_0^2}\, .
\label{p9}
\end{equation}
Since the angle with respect to the beam axis has unique meaning only up to
a value of $\pi / 2$, the paraxial approximation implies that $\theta_0 \ll 1$, 
and consequently that $z_0 \gg w_0 \gg \lambda$.

\begin{figure}[htp]  
\begin{center}
\vspace{0.1in}
\includegraphics[width=3in]{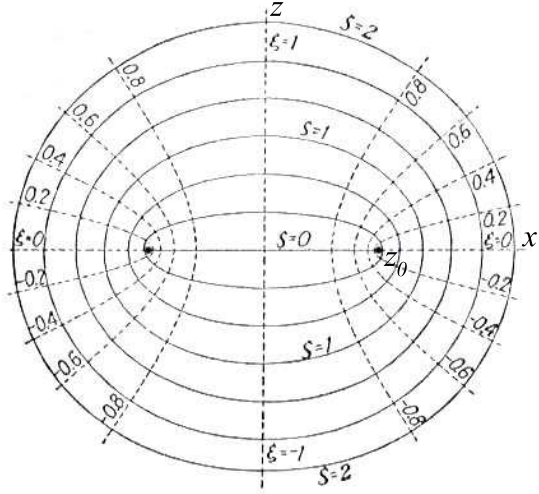}
\parbox{5.5in} 
{\caption[ Short caption for table of contents ]
{\label{fig1} The $x$-$z$ plane of an oblate spheroidal coordinate
system $(\zzeta,\xi,\phi)$ based on hyperboloids and ellipsoids of 
revolution about the $z$ axis, with foci at $(x,z) = (\pm z_0, 0)$.
The coordinates have ranges $0 \leq \zzeta < \infty$, $-1 \leq \xi \leq 1$,
and $0 \leq \phi \leq 2 \pi)$. 
}}
\end{center}
\end{figure}

The question arises whether there are any ``exact'' solutions to the
free-space wave equation
\begin{equation}
\nabla^2 \psi - {1 \over c^2} {\partial^2 \psi \over \partial t^2} = 0,
\label{p0}
\end{equation}
for which the paraxial wavefunctions are a suitable approximation.
For monochromatic waves, it suffices to seek ``exact'' solutions to 
the Helmholtz wave equation,
\begin{equation}
\nabla^2 \psi + k^2 \psi = 0.
\label{p1}
\end{equation}
This equation is known to be separable in 11 coordinate systems 
\cite{Eisenhart,Morse},
of which oblate spheroidal coordinates are well matched to the geometry
of laser beams, as shown in Fig.~\ref{fig1}.

``Exact'' solutions to the Helmholtz equation in oblate spheroidal 
coordinates were developed in the 1930's, and are summarized in
\cite{Stratton,Flammer,Abramowitz}.  These solutions are, however, rather
intricate and were almost forgotten at the time of the invention of
the laser in 1960 \cite{comment}.

This problem does not explore the ``exact'' solutions, but rather asks you
to develop a systematic set of approximate solutions to the Helmholtz
equation in oblate spheroidal coordinates, which will turn out to be one
representation of paraxial Gaussian laser beams.

The relation between rectangular coordinates $(x,y,z)$ and
oblate spheroidal coordinates\footnote{Oblate spheroidal coordinates
are sometimes written with $\zzeta = \sinh u$ and $\xi = \cos v$ or $\sin v$.} 
$(\zzeta,\xi,\phi)$ is
\begin{eqnarray}
x & = & z_0 \sqrt{1 + \zzeta^2} \sqrt{1 - \xi^2} \cos\phi,
\label{p2} \\
y & = & z_0 \sqrt{1 + \zzeta^2} \sqrt{1 - \xi^2} \sin\phi,
\label{p3} \\
z & = & z_0 \zzeta \xi,
\label{p4}
\end{eqnarray}
where the length $z_0$ is the distance from the origin to one of the
foci of the ellipses and hyperbolae whose surfaces of revolution about
the $z$ axis are surfaces of constant $\zzeta$ and $\xi$.  Coordinate
$\phi$ is the usual azimuthal angle measured in the $x$-$y$ plane. 
For large $\zzeta$, the oblate
spheroidal coordinates are essentially identical to spherical coordinates
$(r,\theta,\phi)$ with the identification $\zzeta = r / z_0$ and $\xi = 
\cos\theta$. 

An obvious consequence of the definitions (\ref{p2})-(\ref{p4}) is that
\begin{equation}
r_\perp = \sqrt{x^2 + y^2} = z_0 \sqrt{1 + \zzeta^2} \sqrt{1 - \xi^2}.
\label{p5}
\end{equation}

It is clear that the oblate spheroidal wave functions will
have the mathematical
restriction that the entire wave crosses the plane $z = 0$ within an iris
of radius $z_0$, the length used in the definitions (\ref{p2})-(\ref{p4})
of the oblate spheroidal coordinates.  In effect, the plane $z = 0$ is
perfectly absorbing except for the iris of radius $z_0$.  

You will find that the length $z_0$ also has the physical significance of
the Rayleigh range, which concept is usually associated with longitudinal
rather than tranverse behavior of the waves.  Since the paraxial approximation
that you will explore is valid only when the beam waist $w_0$ is small 
compared to the Rayleigh range, \ie, when $w_0 \ll z_0$,
the paraxial wave functions are not accurate descriptions of waves of 
extremely short focal length, even though they will be formally defined 
for any value of $w_0$. 

The wave equation (\ref{p1}) is separable in oblate spheroidal coordinates,
with the form
\begin{equation}
{\partial \over \partial \zzeta} (1 + \zzeta^2) {\partial \psi \over \partial \zzeta}
+ {\partial \over \partial \xi} (1 - \xi^2) {\partial \psi \over \partial \xi}
+ {\zzeta^2 + \xi^2 \over (1 + \zzeta^2) (1 - \xi^2)}
{\partial^2 \psi \over \partial \phi^2}
 + k^2 z_0^2 (\zzeta^2 + \xi^2) \psi = 0.
\label{p6}
\end{equation}

It is helpful to express the wave functions in radial and transverse coordinates
that are scaled by the Rayleigh range $z_0$ and by the diffraction angle
$\theta_0$, respectively.
The oblate spheroidal coordinate $\zzeta$ already has this desirable property
for large values.  However, the coordinate $\xi$ is usefully replaced by
\begin{equation}
\sigma = {1 - \xi^2 \over \theta_0^2}
= {z_0^2 \over w_0^2} (1 - \xi^2)
= {k z_0 \over 2} (1 - \xi^2)
= {r_\perp^2 \over w_0^2 (1 + \zzeta^2)}\, ,
\label{p6a}
\end{equation}
which obeys $\sigma \approx (\theta / \theta_0)^2$ for large $r$ and small $\theta$,
and $\sigma \approx (r_\perp / w_0)^2$ near the beam waist where $\zzeta \approx 0$.

To replace $\xi$ by $\sigma$ in the Helmholtz equation (\ref{p6}), note that
$2 \xi d\xi = - \theta_0^2 d\sigma$.  In the paraxial approximation, $\xi \approx
1$ (which implies that your solution will be restricted to waves in the
hemisphere $z \geq 0$), you may to suppose that 
\begin{equation}
d \xi \approx - {\theta_0^2 \over 2} d\sigma.
\label{p6b}
\end{equation}

Find an orthogonal set of waves,
\begin{equation}
\psi_n^m = Z_n^m(\zzeta) S_n^m(\sigma) e^{\pm i m \phi},
\label{p7}
\end{equation}
which satisfy the Helmholtz equation in the paraxial approximation.
You may anticipate that the ``angular'' functions $S_n^m(\sigma)$ are
modulated Gaussians, containing a factor $\sigma^{m / 2} e^{-\sigma}$.
The ``radial'' functions $Z_n^m$ are modulated spherical waves in the far
zone, with a leading factor $e^{i k r}$, and it suffices to keep terms in 
the remaining factor that are lowest order in the small quantity $\theta_0$.

Vector electromagnetic waves ${\bf E} = {\bf E}({\bf x}) e^{-i \omega t}$
 and ${\bf B} = {\bf B}({\bf x}) e^{-i \omega t}$ that satisfy Maxwell's
equations in free space can be generated from the scalar wave functions $\psi_n^m$
by supposing the vector
potential {\bf A} has Cartesian components (for which $(\nabla^2 {\bf A})_j =
\nabla^2 A_j$ \cite{Morse}) given by one or more of the scalar 
waves  $\psi_n^m e^{-i \omega t}$.
For these waves, the fourth Maxwell equation in free space is
$c \nabla \times {\bf B} = \partial {\bf E} / \partial t = - i k {\bf E}$
(Gaussian units), so both fields {\bf E} and {\bf B} can de derived 
from the vector potential {\bf A} according to,
 \begin{equation}
{\bf E} = {i \over k} \nabla \times {\bf B}
=  i k {\bf A} + {i \over k} \nabla (\nabla \cdot {\bf A}),
\qquad {\bf B} = \nabla \times {\bf A},
\label{p11}
\end{equation}
since the vector potential obeys the Helmholtz equation (\ref{p1}).

Calculate the ratio of the angular momentum density of the wave in the far zone 
to its energy density
 to show that quanta of these waves (photons with intrinsic spin $S = 1$) 
carry orbital angular momentum in addition to the intrinsic spin.  Show also
that lines of the Poynting flux form spirals on a cone in the far zone.

\section{Solution}

\subsection{The Paraxial Gaussian-Laguerre Wave Functions}

Using the approximation (\ref{p6b}) when replacing variable $\xi$ by
$\sigma$, the Helmholtz equation (\ref{p6}) becomes 
\begin{equation}
{\partial \over \partial \zzeta} (1 + \zzeta^2) {\partial \psi \over \partial \zzeta}
+ {4 \over \theta_0^2} {\partial \over \partial \sigma} \sigma 
{\partial \psi \over \partial \sigma}
+ {1 + \zzeta^2 - \theta_0^2 \sigma \over (1 + \zzeta^2) \theta_0^2 \sigma}
{\partial^2 \psi \over \partial \phi^2}
 + {4 \over \theta_0^4} (1 + \zzeta^2 - \theta_0^2 \sigma) \psi = 0.
\label{s1}
\end{equation}
This equation admits separated solutions of the form (\ref{p7}) for any integer
$m$.  Inserting this in eq.~(\ref{s1}) and dividing by $\psi$, we find
\begin{equation}
{1 \over Z} {\partial \over \partial \zzeta} (1 + \zzeta^2) 
{\partial Z \over \partial \zzeta}
+ {4 \over \theta_0^2 S} {\partial \over \partial \sigma} \sigma 
{\partial S \over \partial \sigma}
- m^2 {1 + \zzeta^2 - \theta_0^2 \sigma \over (1 + \zzeta^2) \theta_0^2 \sigma}
 + {4 \over \theta_0^4} (1 + \zzeta^2) - {4 \sigma \over \theta_0^2} = 0.
\label{s2}
\end{equation}
The functions $Z$ and $S$ will be the same for integers $m$ and $-m$, so
henceforth we consider $m$ to be non-negative, and write the azimuthal functions
as $e^{\pm i m \phi}$.
With  $\lambda_m$ as the second separation constant, the
$\zzeta$ and $\sigma$ differential equations are
\begin{eqnarray}
{d \over d \zzeta} (1 + \zzeta^2) {d Z \over d \zzeta}
& = & \left( \lambda_m - {4 \over \theta_0^4} (1 + \zzeta^2) 
- {m^2 \over 1 + \zzeta^2} \right) Z,
\label{s3} \\
{d \over d \sigma} \sigma {d S \over d \sigma}
& = & - {\theta_0^2 \over 4} \left( \lambda_m - {4 \sigma \over \theta_0^2} 
- {m^2 \over \theta_0^2 \sigma} \right) S.
\label{s4}
\end{eqnarray}

The hint is that the wave functions have Gaussian transverse dependence,
which implies that the ``angular'' function $S(\sigma)$ contains a factor
$e^{-\sigma} = e^{-r_\perp^2 / w_0^2 (1 + \zzzeta^2)}$.  We therefore write
$S = e^{-\sigma}T$, and eq.~(\ref{s4}) becomes
\begin{equation}
\sigma {d^2 T \over d \sigma^2} + (1 - 2 \sigma) {d T \over d \sigma}
+  \left( 1 + {\theta_0^2 \over 4} \lambda_m - {m^2 \over 4 \sigma} \right) T
= 0.
\label{s5}
\end{equation}
The function $T(\sigma)$ cannot be represented as a polynomial, but (like the
radial Shr\"odinger equation) this can be accomplished after a factor of
$\sigma^{m / 2}$ is extracted.  That is, we write $T = \sigma^{m/2} L$, or
$S = \sigma^{m / 2} e^{- \sigma} L$, so that eq.~(\ref{s5}) becomes
\begin{equation}
\sigma {d^2 L \over d \sigma^2} + (m + 1 - 2 \sigma) {d L \over d \sigma}
+  \nu L = 0,
\label{s6}
\end{equation}
where
\begin{equation}
\nu = {\theta_0^2 \over 4} \lambda_m - m - 1.
\label{s7}
\end{equation}
If $\nu = 2 n$ for integer $n \geq 0$, this is the differential equation for 
generalized Laguerre polynomials $L_n^m(2\sigma)$ \cite{Laguerre}, where
\begin{equation}
L_n^m(x) = m! n! \sum_{k=0}^n {(-1)^k x^k \over (m+k)! (n-k)! k!}
= 1 - {n x \over m + 1} + {n (n - 1) x^2 \over 2 (m + 1) (m + 2)} - ...
\label{s8}
\end{equation}
By direct calculation from eq.~(\ref{s6}) with $\nu = 2 n$, we readily
verify that the low-order solutions are
\begin{equation}
L_0^m = 1, \qquad L_1^m(2\sigma) = 1 - {2 \sigma \over m + 1}\, ,
\qquad L_2^m(2\sigma) = 1 - {4 \sigma \over m + 1} 
+ {4 \sigma^2 \over (m + 1)(m + 2)}\, .
\label{s9}
\end{equation}

The Laguerre polynomials are normalized to 1 at $x = 0$, and obey the 
orthogonality relation
\begin{equation}
\int_0^\infty L_n^m(x) L_{n'}^m(x) x^m e^{-x}dz 
= {(m!)^2 n! \over (m+n)!} \delta_{nn'}.
\label{s10}
\end{equation}

The ``angular'' functions $S_n^m(\sigma)$ are thus given by
\begin{equation}
S_n^m(\sigma) = \sigma^{m/2} e^{-\sigma} L_n^m(2\sigma),
\label{s11}
\end{equation}
which obey the 
orthogonality relation
\begin{equation}
\int_0^\infty S_n^m(\sigma) S_{n'}^m(\sigma) d\sigma
= {1 \over 2^{m+1}} \int_0^\infty L_n^m(x) L_{n'}^m(x) x^m e^{-x}dz 
= {(m!)^2 n! \over (m+n)! 2^{m+1}} \delta_{nn'}.
\label{s12}
\end{equation}
In the present application, $0 \leq \sigma \leq 1 / \theta_0^2$, 
on which interval the functions $S_n^m$ are only approximately 
orthogonal.  Because of the exponential damping of the $S_n^m$, their
orthogonality is nearly exact for $\theta_0 \lsim 1/2$.

We now turn to the ``radial'' functions $Z_n^m(\zzeta)$ which obey the
differential equation (\ref{s3}) with separation constant $\lambda_m$
given by
\begin{equation}
\lambda_m = {4 \over \theta_0^2} (2n + m + 1),           
\label{s13}
\end{equation}
using eq.~(\ref{s7}) with $\nu = 2 n$.  For large $r$ the radial functions
are essentially spherical waves, and hence have leading dependence $e^{i k r}$.
For small polar angles, where $\xi \approx 1$, the relation (\ref{p4})
implies that $r \approx z_0 \zzeta$, and $k r \approx k z_0 \zzeta = 2 \zzeta /
\theta_0^2$, recalling eq.~(\ref{p9}).  Hence we expect the radial functions
to have the form\footnote{It turns out not to be useful to extract a factor
$e^{i k r}/ r$ from the radial functions, although these functions will
have this form asymptotically.}
\begin{equation}
Z(\zzeta) = e^{2 i \zzzeta / \theta_0^2} F(\zzeta).           
\label{s14}
\end{equation}
Inserting this in eq.~(\ref{s3}), we find that function $F$ obeys the
second-order differential equation
\begin{equation}
\left( 1 + \zzeta^2 \right) \left( {d^2 F \over d \zzeta^2} 
+ {4 i \over \theta_0^2} {d F \over d \zzeta} \right) +
2 \zzeta \left(  {d F \over d \zzeta} + {2 i F \over \theta_0^2} \right)
= \left( {4 \over \theta_0^2} (2n + m + 1) - {m^2 \over 1 + \zzeta^2}
\right) F.     
\label{s15}
\end{equation}

In the paraxial approximation, $\theta_0$ is small, so we keep only those terms
in eq.~(\ref{s15}) that vary as $1 / \theta_0^2$, which yields the first-order
differential equation,
\begin{equation}
\left( 1 + \zzeta^2 \right) {d F \over d \zzeta} = 
- \left( \zzeta + i (2n + m + 1) \right) F.     
\label{s16}
\end{equation}
For $m = n = 0$ we write $F_0^0 = f$, in which case eq.~(\ref{s16}) reduces to
\begin{equation}
\left( 1 + \zzeta^2 \right) {d f \over d \zzeta} 
= \left( \zzeta + i \right) \left( \zzeta - i \right) {d f \over d \zzeta} 
= - \left( \zzeta + i \right) f,     
\label{s17}
\end{equation}
or
\begin{equation}
{d f \over f} = - {d \zzeta \over \zzeta - i}\, .     
\label{s18}
\end{equation}
This integrates to $\ln f = \ln C - \ln \left( \zzeta - i \right)$.
We define $f(0) = 1$, so that $C = - i$ and
\begin{equation}
f = {1 \over 1 + i \zzeta} = {1 - i \zzeta \over 1 + \zzeta^2}
= {e^{-i \tan^{-1} \zzzeta} \over \sqrt{1 + \zzeta^2}}\, .    
\label{s19}
\end{equation}
At large $\zzeta$, $f \approx 1 / \zzeta \propto 1 / r$, as expected in
the far zone for waves that have a narrow waist at $z = 0$.  Indeed, we
expect that $F_n^m \propto 1 / \zzeta$ at large $\zzeta$ for all $m$ and $n$.
This suggests that $F_n^m$ differs from $f$ by only a phase change.  A
suitable form is
\begin{equation}
F_n^m = {e^{-i a_{m,n} \tan^{-1} \zzzeta} \over \sqrt{1 + \zzeta^2}}
= {\left( e^{-i \tan^{-1} \zzzeta} \right)^{a_{m,n}} \over \sqrt{1 + \zzeta^2}}
= {1 \over \sqrt{1 + \zzeta^2}}\left( {1 - i \zzeta \over \sqrt{1 + \zzeta^2}} \right)^{a_{m,n}}
= {\left( 1 - i \zzeta\right)^{a_{m,n}} \over 
\left( 1 + \zzeta^2 \right)^{(1 + a_{m,n}) / 2}}\, .    
\label{s20}
\end{equation}
Inserting this hypothesis in the differential equation (\ref{s16}), we find
that it is satisfied provided
\begin{equation}
a_{m,n} = 2n + m + 1.    
\label{s21}
\end{equation}
Thus, the radial function is
\begin{equation}
Z_n^m(\zzeta) = e^{i k z_0 \zzzeta} F_n^m 
= {e^{i [k z_0 \zzzeta - (2n + m + 1)  \tan^{-1} \zzzeta]} \over 
\sqrt{1 + \zzeta^2}}\, ,    
\label{s22}
\end{equation}
and the paraxial Gaussian-Laguerre wave functions are
\begin{equation}
\psi_n^m(\sigma,\phi,\zzeta,t) = Z_n^m S_n^m e^{\pm i m \phi} e^{- i \omega t}
= {\sigma^{m / 2} e^{-\sigma} L_n^m(2\sigma)  
e^{i [k z_0 \zzzeta - \omega t - (2n + m + 1) \tan^{-1} \zzzeta \pm m \phi]} 
\over \sqrt{1 + \zzeta^2}}\, .    
\label{s23}
\end{equation}

The factor $e^{-i (2n + m + 1) \tan^{-1} \zzzeta}$ in the 
wave functions implies a phase shift of $(2n + m + 1) \pi /2$ between the
focal plane and the far field, as first noticed by Guoy \cite{Guoy} for whom
this effect is named.  Even the lowest mode, with $m = n = 0$, has a
Guoy phase shift of $\pi / 2$.  This phase shift is an essential
difference between a plane wave and a wave that is asymptotically plane
but which has emerged from a focal region.  The existence of this phase
shift can be deduced from an elementary argument that applies Faraday's
law to wave propagation through an aperture \cite{Faraday}, as 
well as by arguments based on the Kirchhoff diffraction integral 
\cite{diff} as were used by Guoy.

It is useful to relate the coordinates $\sigma$ and $\zzeta$ to those of a cylindrical coordinate system $(r_\perp,\phi,z)$, in the paraxial 
approximation that $\xi \approx 1$.  For this, we recall from eqs.~(\ref{p4}),
(\ref{p5}) and (\ref{p6a}) that
\begin{equation}
\xi = 1 - \theta_0^2 \sigma \approx 1,
\label{s24}
\end{equation}
so
\begin{equation}
\zzeta^2 = {z^2 \over z_0^2 \xi^2} \approx
{z^2 \over z_0^2} (1 + \theta_0^2 \sigma),
\label{s25}
\end{equation}
and hence,
\begin{equation}
r_\perp^2 = w_0^2 \sigma (1 + \zzeta^2) 
\approx w_0^2 \sigma \left(1 + {z^2 \over z_0^2} \right)
+ \theta_0^4 \sigma z^2
\approx w_0^2 \sigma \left(1 + {z^2 \over z_0^2} \right),
\label{s26}
\end{equation}
where we neglect terms in $\theta_0^4$ in the lowest-order paraxial
approximation.  Then,
\begin{equation}
\sigma \approx {r_\perp^2 \over w_0^2 (1 + z^2/z_0^2)}\, , 
\label{s27}
\end{equation}
and
\begin{equation}
\zzeta \approx {z \over z_0} \left(1 + {\theta_0^2 \sigma \over 2} \right)
\approx {z \over z_0} \left(1 + {\theta_0^2 r_\perp^2 \over 2 w_0^2 (1 + z^2/z_0^2)} \right)
= {z \over z_0} \left(1 + {r_\perp^2 \over 2 (z^2 + z_0^2)}\right).
\label{s28}
\end{equation}
For large $z$ eq.~(\ref{s28}) becomes
\begin{equation}
\zzeta \approx {z \over z_0} \left(1 + {r_\perp^2 \over 2 z^2}\right)
\approx {z \over z_0} \sqrt{1 + {r_\perp^2 \over z^2}}
= {r \over z_0}\, ,
\label{s29}
\end{equation}
as expected.  That is, the factor $e^{i (k z_0 \zzzeta - \omega t)}$ in the 
wave functions (\ref{s23}) implies that they are spherical waves in the far zone.

The characteristic transverse extent of the waves at position $z$ is sometimes
called $w(z)$.  From eq.~(\ref{s27}) we see that the Gaussian behavior 
$e^{-\sigma}$ of the angular functions implies that
\begin{equation}
w(z) = w_0 \sqrt{1 + {z^ 2\over z_0^2}}\, .
\label{s30}
\end{equation}

The paraxial approximation is often taken to mean that variable $\zzeta$ is
simply $z/z_0$ everywhere in eq.~(\ref{s23}) except in the phase factor
$e^{i k z_0 \zzzeta}$, where the form (\ref{s28}) is required so that the
waves are spherical in the far zone.  In this convention, we can write
\begin{equation}
\psi_n^m(r_\perp,\phi,z,t) 
= {\sigma^{m/2} e^{-\sigma} L_n^m(2\sigma) 
e^{i \{ k z[1 + r_\perp^2 / 2 (z^2 + z_0^2)] - \omega t 
- (2n + m + 1) \tan^{-1} (z/z_0) \pm m \phi \} } 
\over 
\sqrt{1 + z^2/z_0^2}}\, .    
\label{s31}
\end{equation}

The wave functions may be written in a slightly more compact form if we
use the scaled coordinates 
\begin{equation}
\rho = {r_\perp \over w_0}\, , \qquad
\zzeta = {z \over z_0}\, , \qquad
\sigma = {\rho^2 \over 1 + \zzeta^2}\, .  
\label{s32}
\end{equation}
Then, the simplest wave function is
\begin{eqnarray}
\psi_0^0(r_\perp,\phi,z,t) 
& = & e^{- \rho^2 / (1 + \zzzeta^2)}   
e^{i k z r_\perp^2 / 2 z_0^2 (1 + \zzzeta^2)}
e^{i (k z - \omega t)}
{e^{-i \tan^{-1} \zzzeta} \over \sqrt{1 + \zzeta^2}}
\nonumber \\
& = & f e^{- \rho^2 (1 - i \zzzeta) / (1 + \zzzeta^2)}   
e^{i (k z - \omega t)}
= f e^{- f \rho^2} e^{i (k z - \omega t)},
\label{s33}
\end{eqnarray}
recalling eq.~(\ref{p9}) and the definition of $f(\zzeta)$ in  eq.~(\ref{s19}).
In this manner the general, paraxial wave function can be written
\begin{equation}
\psi_n^m(r_\perp,\phi,z,t) 
= f^{m + 1} \rho^m e^{-f \rho^2} L_n^m(2\sigma) 
e^{i ( k z - \omega t \pm m \phi - 2n \tan^{-1} \zzzeta )}.    
\label{s34}
\end{equation}

It is noteworthy that although our solution began with the hypothesis of
separation of variables in oblate spheroidal coordinates, we have found
wave functions that contain the factors $e^{-f \rho^2}$ and $L_n^m(\sigma)$
that are nonseparable functions of $r_\perp$ and $z$ in cylindrical coordinates.

The wave functions found above are for a pure frequency $\omega$.  In practice
one is often interested in pulses of characteristic width $\tau$ whose
frequency spectrum is centered on frequency $\omega$.  In this case
we can replace the factor $e^{i(k z - \omega t)}$ in the wave function by
$g(\varphi) e^{i \varphi}$, where the phase is $\varphi = k z - \omega t$,
and still satisfy the wave equation (\ref{p0})
provided that the modulation factor $g$ obeys \cite{axicon}
\begin{equation}
\abs{g' \over g} \ll 1.   
\label{s35}
\end{equation}
An important example of a pulse shape that satisfies eq.~(\ref{s35}) is
\begin{equation}
g(\varphi) = {\rm sech} {\varphi \over \omega \tau}\, , 
\label{s36}
\end{equation}
so long as $\omega \tau \gg 1$, \ie, so long as the pulse is longer that
a few periods of the carrier wave.  Perhaps surprisingly, a Gaussian
temporal profile is not consistent with condition (\ref{s35}).  Hence, a
``Gaussian beam" can have a Gaussian transverse profile, but not a Gaussian longitudinal profile as well.

\subsection{Electric and Magnetic Fields of Gaussian Beams}

The scalar wave functions (\ref{s34}) can be used to generate vector
electromagnetic fields that satisfy Maxwell's equations.  For this,
we use eqs.~(\ref{p11}) with a vector potential {\bf A} whose Cartesian
components are one or more of the functions (\ref{s34}).

If we wish to express the electromagnetic fields in cylindrical coordinates,
then we immediately obtain one family of fields from the vector potential
\begin{equation}
A_x = A_y = A_\perp = A_\phi = 0, \qquad
A_z = {E_0 \over k \theta_0} \psi_n^m(r_\perp,\phi,z,t).
\label{s37}
\end{equation}
The resulting magnetic field has no $z$ component, so we may call these
transverse magnetic (TM) waves.  If index $m = 0$ then {\bf A} has no
$\phi$ dependence, and the magnetic field has no radial component; the
magnetic field lines are circles about the $z$ axis.

The lowest-order TM mode, corresponding to indices $m = n = 0$, has
field components
\begin{eqnarray}
E_\perp & = &
E_0 \rho f^2 e^{-f \rho^2} g e^{i \varphi} + {\cal O}(\theta_0^2),
\nonumber \\
E_\phi & = & 0,
\nonumber \\
E_z & = & i \theta_0 E_0 f^2 (1 - f \rho^2) e^{-f \rho^2} g 
e^{i \varphi} + {\cal O}(\theta_0^3).
\nonumber \\
B_\perp & = & 0,
\nonumber \\
B_\phi & = & E_\perp,
\nonumber \\
B_z & = & 0,
\label{s38}
\end{eqnarray}
as apparently first deduced in \cite{Davis81}.  This is a so-called axicon
mode \cite{axicon}, in which the electric field is dominantly radial, which
component necessarily vanishes along the beam axis.  In the far zone
the beam intensity is largest on a cone of half angle $\theta_0$ and is
very small on the axis itself; the beam appears to have a hole in the center.

\begin{figure}[htp]  
\begin{center}
\includegraphics[width=4in, angle=0, clip]{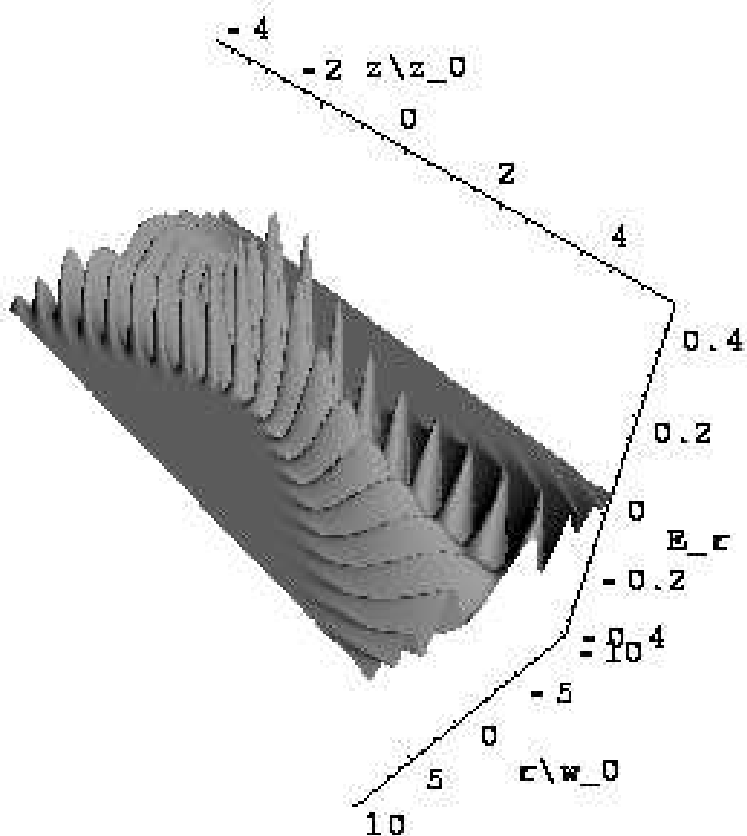} 
\parbox{5.5in} 
{\caption[ Short caption for table of contents ]
{\label{era} The electric field $E_r(r_\perp,0,z)$ of the TM$_0^0$
 axicon Gaussian
beam with diffraction angle $\theta_0 = 0.45$, according to eq.~(\ref{s38}).
}}
\end{center}
\end{figure}

\begin{figure}[htp]  
\begin{center}
\includegraphics[width=4in, angle=0, clip]{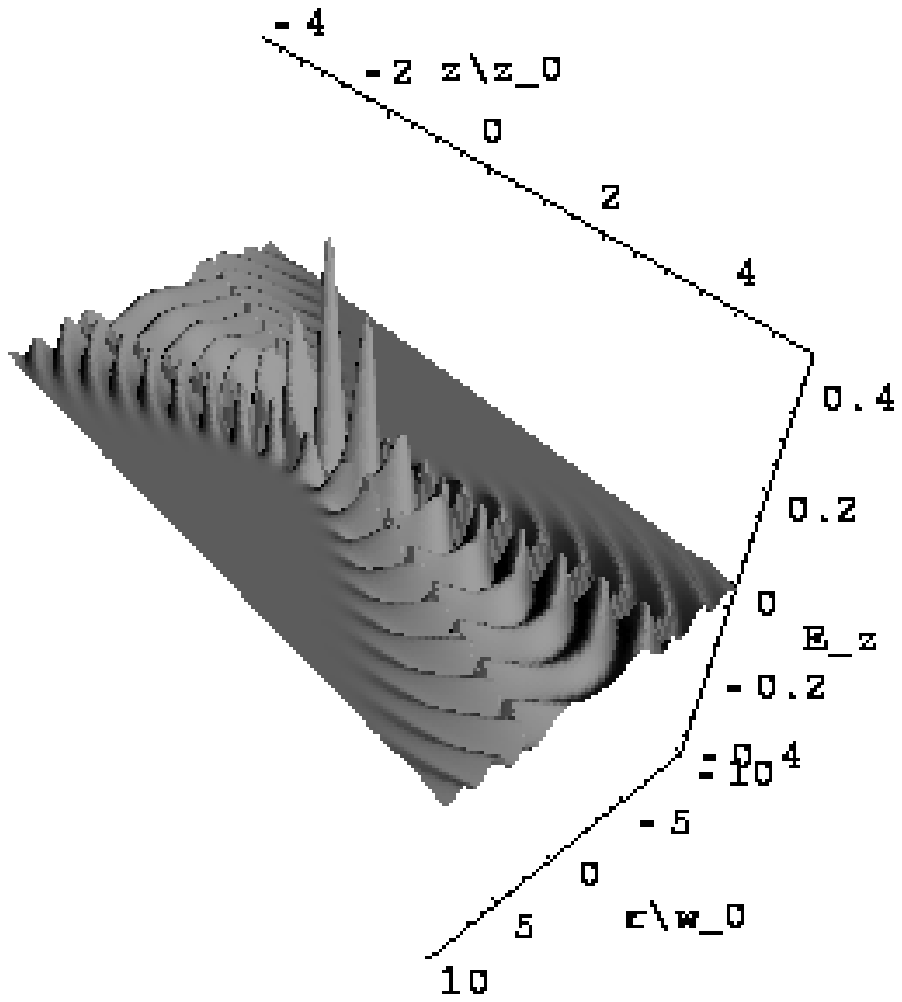} 
\parbox{5.5in} 
{\caption[ Short caption for table of contents ]
{\label{eza} The electric field $E_z(r_\perp,0,z)$ of the TM$_0^0$
 axicon Gaussian
beam with diffraction angle $\theta_0 = 0.45$, according to eq.~(\ref{s38}).
}}
\end{center}
\end{figure}

\begin{figure}[htp]  
\begin{center}
\includegraphics*[width=1.4in]{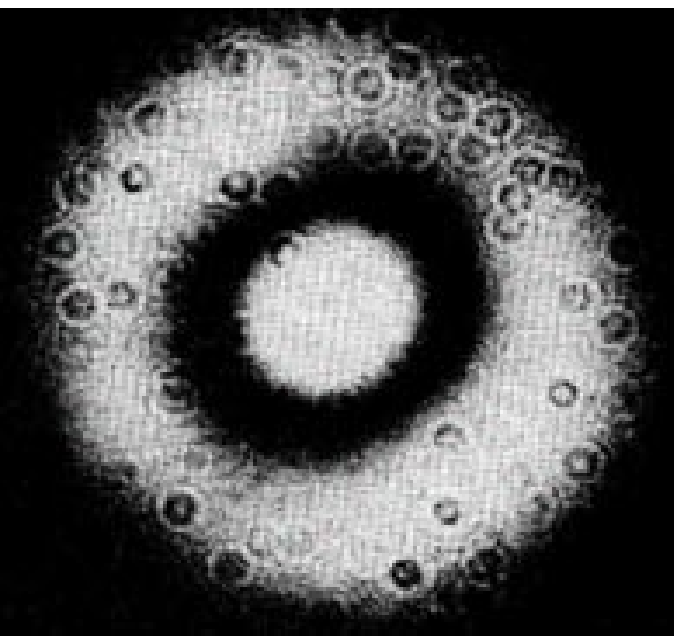}
\includegraphics*[width=1.4in]{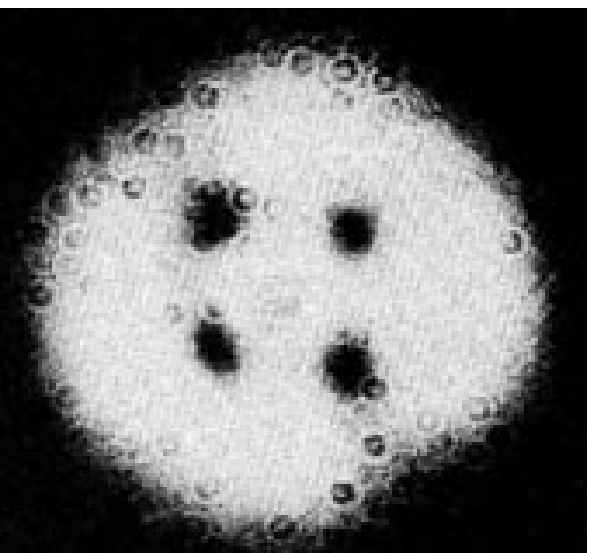}
\includegraphics*[width=1.45in]{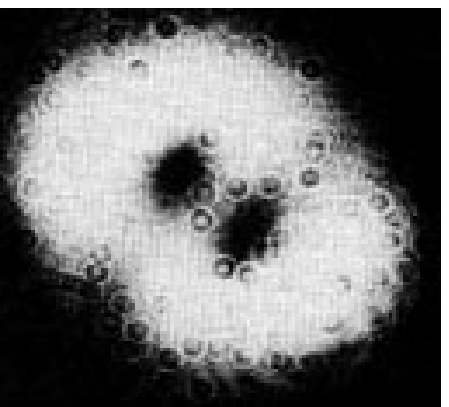}
\includegraphics*[width=1.5in]{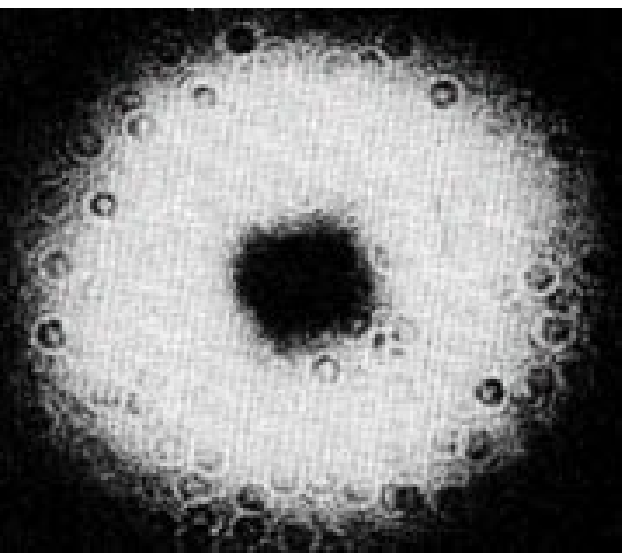} 
\parbox{5.5in} 
{\caption[ Short caption for table of contents ]
{\label{bram} Photographs of Gaussian-Laguerre laser beams with $2n + m = 2$.
From \cite{Brambilla}.
}}
\end{center}
\end{figure}

The radial and longitudinal electric field of the TM$_0^0$ mode are illustrated
in Figs.~\ref{era} and \ref{eza}.  Photographs of Gaussian-Laguerre laser
modes from \cite{Brambilla} are shown in Fig.~\ref{bram}.

As is well known, corresponding to each TM wave solution to Maxwell's
equations in free space, there is a TE (transverse electric) mode obtained
by the duality transformation 
\begin{equation}
{\bf E}_{\rm TE} = {\bf B}_{\rm TM}, \qquad
{\bf B}_{\rm TE} = - {\bf E}_{\rm TM}.
\label{s39}
\end{equation}
Since we are considering waves in free space where $\nabla \cdot {\bf E} = 0$,
the electric field could also be deduced from a vector potential, and the
magnetic field from the electric field, according to the dual of eq.~(\ref{p11}),
\begin{equation}
{\bf E} = \nabla \times {\bf A}, \qquad
{\bf B} = - {i \over k} \nabla \times {\bf E}.
\label{s40}
\end{equation}
Then, the TE modes can be obtained by use of the vector potential (\ref{s37})
in eq.~(\ref{s40}).

The TM Gaussian-Laguerre modes emphasize radial polarization of the electric
field, and the TE modes emphasize circular polarization.  In many physical
applications, linear polarization is more natural, for which the modes are
well-described by Gaussian-Hermite wave functions 
\cite{Goubau,Boyd,Kogelnik,Siegman}.  Formal transformations between the
Gaussian-Hermite wave functions and the Gaussian-Laguerre functions have 
been described in \cite{Abram}.

\subsection{Energy, Momentum and Angular Momentum in the Far Zone}

The electromagnetic field energy density,
\begin{equation}
u = {E^2 + B^2 \over 8 \pi}\, ,
\label{s41}
\end{equation}
the field momentum density,
\begin{equation}
{\bf p} = {{\bf E} \times {\bf B} \over 4 \pi c}\, ,
\label{s42}
\end{equation}
and the field angular momentum density,
\begin{equation}
{\bf l} = {\bf r} \times {\bf p},
\label{s43}
\end{equation}
are the same for a TM Gaussian-Laguerre mode and the TE mode related to it by
the duality transformation (\ref{s39}).

We consider the energy, momentum and angular momentum for TM waves
in the far zone, where
$\zzeta \approx r / z_0 \gg 1$, and $r_\perp  \approx r \theta \ll r$
in terms of spherical coordinates $(r,\theta,\phi)$.  Then
the waves are nearly spherical, and so have a phase factor $e^{i k r}$ that
implies the electric field is related to the magnetic field by
\begin{equation}
{\bf E} = {i \over k} \nabla \times {\bf B} \approx {\bf B} \times \hat{\bf r},
\label{s44}
\end{equation}
so that $E^2 = B^2$.  The time-averaged densities can therefore be written
\begin{equation}
\ave{u} = {\abs{B}^2 \over 8 \pi}\, ,
\label{s45}
\end{equation}
\begin{equation}
\ave{{\bf p}} = {Re[({\bf B} \times \hat{\bf r}) \times {\bf B}^\star] \over 8 \pi c}
= {\abs{B}^2 \hat{\bf r} - Re[({\bf B} \cdot \hat{\bf r}) {\bf B}^\star] 
\over 8 \pi c}
\, ,
\label{s46}
\end{equation}
and
\begin{equation}
\ave{{\bf l}} = - {Re [({\bf B} \cdot \hat{\bf r}) ({\bf r} \times {\bf B}^\star)] 
\over 8 \pi c}\, .
\label{s47}
\end{equation}

The TM waves are derived from the vector potential (\ref{s37}) whose only
nonzero component is $A_z$.  Then, the magnetic field components  in
cylindrical coordinates are
\begin{equation}
B_\perp = {1 \over r_\perp} {\partial A_z \over \partial \phi}
= {\pm i m A_z \over r_\perp},
\qquad
B_\phi = - {\partial A_z \over \partial r_\perp}
= - {1 \over w_0} {\partial A_z \over \partial \rho},
\qquad
B_z = 0,
\label{s48}
\end{equation}
where $\rho = r_\perp / w_0$.
The radius vector {\bf r} has cylindrical components $(r_\perp,0,z)$, so 
\begin{equation}
{\bf B} \cdot \hat{\bf r} = {B_\perp r_\perp \over r}, 
\qquad \mbox{and} \qquad
{\bf r} \times {\bf B}^\star = - \hat{\bf r}_\perp z B_\phi^\star +
\hat\phi z B_\perp^\star + \hat{\bf z} r_\perp B_\phi^\star.
\label{s49}
\end{equation}
Only the $z$ component of the angular momentum can be nonzero for the beam
as a whole, so we calculate
\begin{eqnarray}
\ave{l_z} & = & - { r_\perp^2 \over 8 \pi c r} Re (B_\perp B_\phi^\star) 
\approx {\pm m \over 8 \pi c w_0} \theta Re \left( i A_z 
{\partial A_z^\star \over \partial \rho}\right)
= {\pm m \over 8 \pi c w_0} {2 \theta \over k w_0 \theta_0} Re \left( i A_z 
{\partial A_z^\star \over \partial \rho}\right)
\nonumber \\
& = & {\pm m \over 4 \pi \omega w_0^2} {\theta \over \theta_0} 
Re \left( i A_z {\partial A_z^\star \over \partial \rho}\right),
\label{s50}
\end{eqnarray}
recalling that $r_\perp / r \approx \theta$
 and $\theta_0 = 2 / k w_0^2$.
The factors of $A_z$ that depend on $r_\perp$ and $\sigma \approx z_0^2 \rho^2
/ r^2$ are
\begin{equation}
A_z \propto \rho^m e^{- f \rho^2} L_n^m(2 \sigma)
\approx \rho^m e^{- f \rho^2} \left( 1 - {2 n z_0^2 \rho^2 \over (m+1) r^2} \right),
\label{s51}
\end{equation}
where in the far zone, $f(\zzeta) \approx - i z_0 / r$.  Thus,
\begin{equation}
{\partial A_z \over \partial \rho} \approx
\left( {m \over \rho^2} + 2 i {z_0 \over r} - {4 n z_0^2 \over (m+1) r^2} \right) 
\rho A_z
\approx 2 i {z_0 \over r} \rho A_z
\approx 2 i {\theta \over \theta_0} A_z,
\label{s52}
\end{equation}
since in the far zone the factor $\rho^m e^{-f\rho^2} \approx 
\rho^m e^{- \rho^2 / \zzzeta^2}$ implies that the
wave functions are large only for $\rho \approx \sqrt{m}\
\zzeta \approx \sqrt{m}\ r / z_0 \gg 1$.
Inserting this in eq.~(\ref{s50}), we find
\begin{equation}
\ave{l_z} \approx 
{\pm m \over 2 \pi \omega w_0^2} {\theta^2 \over \theta_0^2} \abs{A_z}^2.
\label{s53}
\end{equation}

To compare with the energy density, we need
\begin{equation}
\abs{B}^2 = \abs{B_\perp}^2 + \abs{B_\phi}^2
= {m^2 \over r_\perp^2} \abs{A_z}^2 + {1 \over w_0^2} 
\abs{\partial A_z \over \partial \rho}^2
\approx {1 \over w_0^2} \left( {m^2 \over \rho^2} + 4 {\theta^2 \over \theta_0^2}
\right) \abs{A_z}^2
\approx {4 \over w_0^2} {\theta^2 \over \theta_0^2} \abs{A_z}^2,
\label{s54}
\end{equation}
since $\rho = \pi r \theta_0 / \lambda \theta \gg \theta_0 / \theta$ in the
far zone.  Using this in eq.~(\ref{s45}), we have
\begin{equation}
\ave{u} \approx {1 \over 2 \pi w_0^2} {\theta^2 \over \theta_0^2} \abs{A_z}^2,
\label{s55}
\end{equation}
and the ratio of angular momentum density to energy density  of a 
Gaussian-Laguerre mode is
\begin{equation}
{\ave{l_z} \over \ave{u}} \approx {\pm m \over \omega}\, ,
\label{s56}
\end{equation}
where the $\pm$ sign corresponds to azimuthal dependence $e^{\pm i m \phi}$.

In a quantum view, the mode contains $N$ photons per unit volume of energy $\hbar
\omega$ each, so the classical result (\ref{s56}) implies that each of these
photons carries orbital angular momentum $\pm m \hbar$.  Since the photons have
intrinsic spin $S = 1$, with $S_z = \pm 1$, we infer that the photons of a 
Gaussian-Laguerre mode carry total angular momentum component $J_z = \pm m \pm 1$.

The angular momentum of Gaussian-Laguerre modes has also been discussed in
\cite{Allen}, in a slightly different approximation.  The first macroscopic
evidence for the angular momentum of light appears to have been given in
\cite{Beth}.

Using the above relations we can evaluate to momentum density (\ref{s46}),
which is proportional to the Poynting vector, and in the far zone we find
\begin{equation}
\ave{{\bf p}} \approx {\ave{u} \over c} \left( \hat{\bf r} \mp
{m \over k r_\perp} \hat\phi \right)
\approx {\ave{u} \over c} \left( \hat{\bf z} + \theta \hat{\bf r}_\perp \mp
{m \over k r_\perp} \hat\phi \right).
\label{s57}
\end{equation}
Since $k r_\perp \gg 1$ in the far zone, the energy flow is largely radial
outward from the focal region.  The small azimuthal component causes the
lines of energy flow to become spirals, which lie on cones of constant
polar angle $\theta$ in the far zone

We can, of course, also deduce the angular momentum density $\ave{l_z} = 
[{\bf r} \times \ave{{\bf p}}]_z$ using eq.~(\ref{s57}).


\begin{thebibliography}{99}
\bibitem{paraxial}
Paraxial Gaussian laser beams were introduced nearly simultaneously
from two different approaches in \cite{Goubau} and \cite{Boyd}.  An
influential review article is \cite{Kogelnik}.  Corrections to the
paraxial approximation were first organized in a power series in
the parameter $\theta_0^2$ in \cite{Lax}.  The understanding that
the scalar, paraxial wave function is best thought of as a component
of the vector potential was first emphasized in \cite{Davis},
with higher-order approximations discussed in \cite{Barton}.  A
textbook with an extensive discussion of Gaussian beams is
\cite{Siegman}.  A recent historical review on the theory and
experiment of laser beam modes is \cite{Siegman2}.  Other
problems on Gaussian laser beams by the author include
\cite{diff}, \cite{axicon} and \cite{bessel}.
\bibitem{Laguerre}
The generalized Laguerre polynomials $L_n^m(x)$ defined in eq.~(\ref{s8}) 
are the Kummer polynomials $M(a,b,z)$ discussed in sec.~13 of \cite{Abramowitz}, 
with $a = - n$, $b = m + 1$ and $z = x$.  The orthogonality relation
(\ref{s10}) is deducible from 22.2.12 of \cite{Abramowitz}, p.~774.
\end{thebibliography}
\end{document}